\def\year{2021}\relax
\documentclass[letterpaper]{article} 
\usepackage{aaai21}  
\usepackage{times}  
\usepackage{helvet} 
\usepackage{courier}  
\usepackage[hyphens]{url}  
\usepackage{graphicx} 
\usepackage{natbib}  
\usepackage{caption} 
\usepackage{booktabs}
\usepackage{makecell}
\usepackage{array}
\usepackage{appendix}
\newcolumntype{L}[1]{>{\raggedright\let\newline\\\arraybackslash\hspace{0pt}}p{#1}}
\newcolumntype{C}[1]{>{\centering\let\newline\\\arraybackslash\hspace{0pt}}p{#1}}
\newcolumntype{R}[1]{>{\raggedleft\let\newline\\\arraybackslash\hspace{0pt}}p{#1}}

\usepackage{lineno}
\usepackage{scalerel,xparse}
\NewDocumentCommand\emojipoop{}{
    \scalerel*{
        \includegraphics{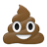}
    }{X}
}
\NewDocumentCommand\emojiclown{}{
    \scalerel*{
        \includegraphics{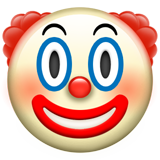}
    }{X}
}
\NewDocumentCommand\emojiwavinghand{}{
    \scalerel*{
        \includegraphics{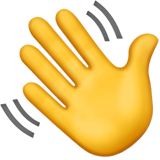}
    }{X}
}
\usepackage{xcolor}
\title{Reconsidering Tweets:\\
Intervening During Tweet Creation Decreases Offensive Content}
\author{
    Matthew Katsaros\textsuperscript{\rm 1},
    Kathy Yang\textsuperscript{\rm 2},
    Lauren Fratamico\textsuperscript{\rm 2}
    \\
}
\affiliations{
    \textsuperscript{\rm 1}Yale Law School\\
    \textsuperscript{\rm 2}Twitter Inc.\\
    matthew.katsaros@yale.edu\\
     \{kathyy, lfratamico\}@twitter.com

}

\begin{document}

\maketitle

\begin{abstract}
The proliferation of harmful and offensive content is a problem that many online platforms face today. One of the most common approaches for moderating offensive content online is via the identification and removal after it has been posted, increasingly assisted by machine learning algorithms. More recently, platforms have begun employing moderation approaches which seek to intervene prior to offensive content being posted. In this paper, we conduct an online randomized controlled experiment on Twitter to evaluate a new intervention that aims to encourage participants to reconsider their offensive content and, ultimately, seeks to reduce the amount of offensive content on the platform. The intervention prompts users who are about to post harmful content with an opportunity to pause and reconsider their Tweet. We find that users in our treatment prompted with this intervention posted 6\% fewer offensive Tweets than non-prompted users in our control. This decrease in the creation of offensive content can be attributed not just to the deletion and revision of prompted Tweets - we also observed a decrease in both the number of offensive Tweets that prompted users create in the future and the number of offensive replies to prompted Tweets. We conclude that interventions allowing users to reconsider their comments can be an effective mechanism for reducing offensive content online.
\end{abstract}

\section{Introduction}
Offensive content, abuse, and harassment online are wide reaching problems: a 2020 Pew survey found that 41\% of US adults reported personally experiencing some form of online harassment \cite{pew2021}. This same survey also found that people experiencing different forms of online harassment, from physical threats to offensive name-calling, has increased since 2014. Wide-reaching and increasing, offensive content online has been associated with various forms of impact ranging from increases in affective polarization \cite{tyagi2020affective} to increases in depressive symptoms in teenagers resulting from cyberbullying \cite{reed2016cyberbullying}.

For as many different types of offensive content that exist online, there are just as many motivations and individual circumstances that result in the creation of that offensive content. Research finds some of the motivations for offensive content can include: retribution \cite{blackwell2018online}, moral outrage \cite{brady2020mad}, or people being in a “hot state” \cite{wang2011regretted}.

In this work, we focus on this last group by offering an intervention at the time of content creation. The intervention encourages people to stop and reconsider their comments, and it provides them with an opportunity to revise their contributions. This approach is one that has recently been adopted by a number of platforms including Instagram\footnote{https://about.fb.com/news/2019/12/our-progress-on-leading-the-fight-against-online-bullying/}, Nextdoor\footnote{https://blog.nextdoor.com/2019/09/18/announcing-our-new-feature-to-promote-kindness-in-neighborhoods/}, YouTube\footnote{https://blog.youtube/news-and-events/make-youtube-more-inclusive-platform/}, Wikipedia\footnote{https://medium.com/jigsaw/helping-authors-understand-toxicity-one-comment-at-a-time-f8b43824cf41}, Pinterest\footnote{https://newsroom.pinterest.com/en/creatorcode}, and TikTok\footnote{https://newsroom.tiktok.com/en-us/new-tools-to-promote-kindness}.

Through an online randomized controlled experiment on Twitter, we test a real-time intervention during Tweet creation that encourages people to reconsider their Tweet. We try to understand how this intervention can be used to not just curb the creation of those Tweets intervened upon, but also change people's behaviors and shape online conversations.

\section{Related Work}
\subsection{Offensive Content Online}
Researchers have used many different names in reference to offensive content including incivility \cite{anderson2018toxic}, hate speech \cite{fortuna2018survey}, and harassment \cite{blackwell2018online} with just as many ways to actually define this content. The lack of a consensus term and definitions for specific problem areas extends beyond the research community as platforms each define these similar issues in different ways \cite{pater2016characterizations}.

The development of Google’s Perspective API, which defines toxic speech as a “rude, disrespectful, or unreasonable comment that is likely to make someone leave a discussion”\footnote{https://www.perspectiveapi.com/}, has given many researchers a common lens for analyzing this problem. Perspective uses machine learning to detect toxic text, and researchers have used Google's Perspective API to assess toxicity of comments on YouTube \cite{obadimu2019youtube}, Reddit \cite{rajadesingan2020reddit}, and the Civil Comments platform \cite{vaidya2020empirical}.

\subsection{Motivations of Offensive Content Online}
Research has been conducted to better understand the circumstances and motivations for offensive content online. While some of this work looks broadly across people's experiences online, other studies investigate a specific motivation behind a common online phenomena.

Through interviews, diary studies, and surveys, \citet{wang2011regretted} provided a broad qualitative understanding of people’s regrets in posting content to Facebook. This research looked at what types of content people regret posting to Facebook and found “negative or offensive comments” to be one of the common themes of regretful postings. When understanding why people regret these postings, common reasons included “venting frustration” or users being in emotional “hot” states which the research describes as “things they posted while in a highly emotional state.” While it’s clear from the research in this space that there are a wide variety of motivations for offensive content online, our experiment focuses on trying to address these circumstances.

\citet{blackwell2018online} conducted online experiments to understand when online harassment might be perceived as justified. This work found that when the target of online harassment has committed some offense, people perceive the harassment to be more deserved and justified (though not any more appropriate). \citet{brady2020mad} propose a psychological model of “moral contagion” on social media platforms associated with some of this offensive content driven by the cycle of: people’s group identity-based motivations to share moral-emotional content; the ability which that moral-emotional content has to grab our attention; and the design of social media platforms to encourage consumption and sharing of moral-emotional content. \citet{seering2017shaping} looked at users' tendencies to mimic others in Twitch chat rooms as a way to explain antisocial behavior within the chat threads. Further still, some have found links between those exhibiting ``dark triad'' personality traits and anti-social behavior online \cite{petit2020associations, bogolyubova2018dark}.

Together, this work demonstrates a wide variety of circumstances and motivations behind various forms of offensive content online. It becomes clear that there will not be a single solution suitable for all of these circumstances; instead, it will be necessary to more precisely tailor interventions and governance approaches to these varying circumstances.

\subsection{Content Moderation}

How to effectively manage the wide spectrum of offensive content online has long been an area of concern for platform owners and researchers alike. \citet{grimmelmann2015virtues} puts forth a taxonomy or “grammar” of moderation through a set of techniques, distinctions, and community characteristics that try to describe various forms of moderation online. In that, he draws a distinction between ex ante and ex post moderation - whether some moderation action takes place before content is allowed to be posted to a community (ex ante) or taking action on content after that content has been posted to the community (ex post).

\subsubsection{Ex Post Moderation}

Currently, one of the predominant mechanisms used online to moderate offensive content is through removal after content has been posted. This applies both for moderation actions taken by large platforms, but also for moderation actions taken in community moderation settings \cite{seering2019moderator}. Increasingly, this approach relies on the use of algorithms and machine learning to identify or help sort and prioritize content \cite{chandrasekharan2017bag, wulczyn2017ex, yin2009detection}. While this can be a tremendously useful approach to manage the massive scale at which many platforms are operating, the use of algorithms for moderation undoubtedly comes with their own risks and issues including false positives \cite{hosseini2017deceiving} and algorithmic bias \cite{vaidya2020empirical, binns2017like}.

Within this context of removing content ex post, some researchers have looked at the role of transparency in the moderation experience. Much of this work suggests that providing more transparency throughout the moderation experience can increase future rule compliance \cite{tyler2019social, jhaver2019does, jhaver2019did}. \citet{jhaver2019does} found that users who had posts removed from Reddit were less likely to have future posts removed when provided explanations. \citet{tyler2019social} similarly found that users who were provided education about platform rules in the week following their post removal were less likely to have their posts removed.

\subsubsection{Ex Ante Moderation}

While acting after something has been posted online seems to be the more common approach, there have no doubt been investigations into trying to take moderation actions before content is posted. Similar to those previously referenced studies looking at the role of transparency during the moderation experience, others have looked at the role transparency has before any moderation takes place. \citet{matias2019preventing} found that posting the rules to the r/Science Reddit discussion board increased rule compliance amongst new members. \citet{jhaver2019did} found that, amongst users on Reddit who had recently had a post removed, those who had previously read the Subreddit's rules were more likely to perceive their post removal as fair.

Some approaches seek to intervene closer to the moment that someone is engaging online, but none have specifically tacked intervening to decrease the presence of offensive content by the poster. \citet{seering2019designing} designed a system using psychologically ``embedded'' CAPTCHAs containing stimuli intended to prime positive emotions and mindsets which significantly increased the positively of sentiment in participant’s posts. \citet{kriplean2012supporting} designed a platform, ConsiderIt, that encouraged healthier debate by prompting users to consider the perspectives of others before posting. \citet{wu2021better} similarly encouraged healthier text interactions by designed an intervention that encourages users to give more-thoughtful and kinder feedback by leveraging the empathy of the feedback giver.

In terms of decreasing the posting of offensive content at time of posting, some work has focused on the roles bystanders can play. \citeauthor{blackwell2017classification}'s work gives promise that designing systems that allow for bystander intervention can reduce the online harassment of an individual, even when the individual had previously committed an offence and dog piling is occurring  \cite{blackwell2017classification, blackwell2018online}. \citet{taylor2019accountability} further found that it was possible to increase the rate of a bystander intervening by designing a system that promoted accountability and empathy.

Finally, as mentioned in the introduction, a number of platforms in the past few years have introduced features to allow users to reconsider their comments before posting when the platform has algorithmically detected the comment may be offensive or otherwise unwanted, though none have shared detailed findings of the efficacy or resulting behavior changes. That is our intention with this work.

\begin{figure}
\centering
\begin{minipage}{.5\textwidth}
  \centering
  \includegraphics[width=0.6\columnwidth]{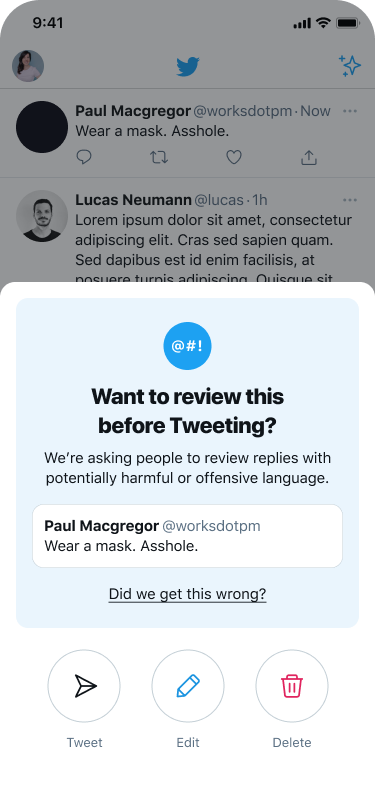}
  \captionof{figure}{Main Intervention Interface}
  \label{nudge_interface}
\end{minipage}
\end{figure}

\section{Hypotheses}
We designed an intervention shown to users as they are about to post an offensive Tweet. With this experiment, we are testing the four following hypotheses:

\begin{itemize}
\item H1: People who are presented an opportunity to pause and consider their offensive Tweet will be less likely to create offensive Tweets.
\item H2: People who are presented an opportunity to pause and consider their offensive Tweet will be less likely to create other similarly offensive Tweets in the future.
\item H3: People who are presented an opportunity to pause and consider their offensive Tweet will be less likely to delete their Tweets after posting, as they are less likely to make regrettable Tweets that are hastily posted. 
\item H4: People who are presented an opportunity to pause and consider their offensive Tweet will receive fewer offensive replies to their Tweets.
\end{itemize}

\section{Intervention Interface Design}
In this experiment, we designed an intervention aimed at interrupting those caught in a ``hot state'' while posting content online. This intervention would allow users to pause and think about their comments to avoid regretful postings. As discussed more below, we hypothesize that intervening during this moment will not only have an impact at the time of intervention (H1), but can help change people's future behaviors (H2 and H3), and shape the conversations these individuals are participating in (H4).

The main interface, displayed in Figure \ref{nudge_interface}, is shown as a user is about to post an offensive reply. Specifically, when a user clicks ``Reply'' after writing an offensive reply to a Tweet, the interface shown is displayed to the user. It displays the text of their proposed post and asks them if they would like to review the post before posting. While this interface is shown, the proposed Tweet is not yet posted to Twitter, and others on the platform cannot see it.

This interface provides the prompted user the options to edit, delete, or post as originally written. If the users clicks ``Tweet'', the Tweet is posted as they originally typed it. Those who click ``Edit'' are allowed to edit the Tweet text inline before sending. The ``Delete'' option cancels the proposed post entirely, taking them back to the Twitter screen they were initially on.

\section{Experimental Design}

\subsection{Eligibility}
To be eligible for the experiment, a user must have made a prompt-eligible Tweet while the experiment was running. The primary criteria for a Tweet being being prompt-eligible were:
\begin{itemize}
    \item Tweet was a reply Tweet in English to another user. Root Tweets were not prompt-eligible, nor were Tweets sent in reply to oneself.
    \item Tweet met or exceeded a preset threshold for offensive content. An internal algorithm was used, which aims to estimate the likelihood that a Tweet's text contains harmful content such as threats or descriptions of physical harm, sexual harassment, or individual attacks. The algorithm is based on BERT (Bidirectional Encoder Representations from Transformers) \citep{devlin2018bert}, a pre-trained language model, and fine-tuned on a set of human-labeled Tweet data.
    \item Any single user was not prompted on more than two Tweets in a single day.%
\end{itemize}

\subsection{Evaluating Offensive Tweets Classifier}

Prior to launching the experiment, we assessed the quality of our prompt-eligibility criteria and the internal algorithm used to both identify Tweets for this experiment and classify Tweets as offensive. Through a third-party platform, prompt-eligible Tweets were sent for annotation by English-speaking individuals with an equal distribution of men and women.

Each annotator reviewed Tweets that met our prompt eligibility criteria. To provide context to the Tweet being reviewed, the annotator was shown both the prompt-eligible reply Tweet (that they were annotating) as well as the Tweet the reply was in response to. The annotator was asked ``Is this Reply Tweet toxic?'' and could select between ``Yes'' and ``No'' labels. The following guideline was provided: ``A toxic Reply Tweet is defined as a rude, disrespectful, or unreasonable comment that is likely to make someone leave a discussion''.

We randomly sampled 2,500 prompt-eligible Tweets made in a one-week period two months prior to the experiment launch date. Each Tweet was labeled by five annotators, and we used a majority vote to obtain the final label for each tweet. Some Tweets were not able to be labeled (e.g., due to the user's privacy settings, or the Tweet being deleted at the time of labeling), and the resulting labeled set included 1,929 Tweets. 95\% of these prompt-eligible Tweets were labeled ``Yes'' in response to the question, ``Is this Reply Tweet toxic?'', and 5\% were labeled ``No''. Inherent to the identification of offensive content is an understanding that the context of language is very important and can be difficult for an outside party to appropriately discern; even Tweets identified as ``Toxic'' by multiple independent annotators could still be viewed as unoffensive by many others (and visa versa). That said, we felt the results of this task to be a satisfactory indication that the criteria for selecting Tweets to prompt in this experiment was sufficiently appropriate. We also felt it was a satisfactory indication that we could accurately classify offensive content. Thus, this is also the algorithm we used to classify offensive content in the Results section.

\subsection{Assignment}
This experiment was set up as a randomized controlled experiment (A/B test), with an equal number of users assigned to control and treatment. Users assigned to control never saw the intervention shown in Figure \ref{nudge_interface} while those assigned to treatment did. 

On the first time a user created a prompt-eligible Tweet during our experiment, they were randomly assigned to the control or treatment condition. Those assigned to treatment would receive the intervention after attempting to send a prompt-eligible reply Tweet, for as many prompt-eligible Tweets as they created for the duration of the experiment (to up two times per day). Users in control would never see the prompt for the duration of the experiment.

Prior to conducting the experiment, a rigorous internal review by Twitter was completed to ensure that it was run ethically and that users’ privacy was preserved. This included a red team exercise \citep{zenko2015red} in addition to review by Twitter's privacy, policy, legal, and research departments. All experiment analysis was performed on de-identified data in accordance with Twitters’s Data Policy \cite{twitter_2021}.

\section{Results}

The experiment began on February 26, 2021 and ended on April 8, 2021, for a total duration of six weeks. During this time, 219,052 users were enrolled in the experiment with the anticipated 50\%/50\% split between test and control (109,551 in control, 109,501 in treatment).
438,943 prompt-eligible Tweets were composed - 226,095 from users assigned to control, 212,848 from users assigned to treatment. The majority of users (63\%) composed only a single prompt-eligible Tweet during the experiment. In Table \ref{dist_of_users_table}, we show the distribution of users by total prompt exposure. 


\begin{table}[h!]
\centering
\caption{Distribution of Users by Total Prompt Exposure during Experiment}
 \resizebox{1.0\columnwidth}{!}{\begin{tabular}{c c} 
 \toprule
 Total Prompt Exposure & Percent of Users in Experiment\\
 \midrule
 1 & 63\%\\ 
 2 & 18\%\\
 3 & 7\%\\
 4 & 4\%\\
 5+ & 8\%\\
 \bottomrule
 \end{tabular}}
\label{dist_of_users_table}
  \small \\
  Note: n = 219,052
\end{table}

\subsection{Hypothesis 1}

To start, we compared the number of offensive replies sent and total replies sent between users assigned to treatment and users assigned to control (summarized in Table \ref{experiment_table}). Offensive replies sent are defined as a Tweet which met or exceeded a preset threshold for offensive content - the same threshold used to identify prompt-eligible Tweets.  We saw a statistically significant decrease in offensive replies sent of 6.4\% (two-tailed p-value of 9.6e-8). This decrease shows strong support of our first hypothesis (H1) - \emph{People who are presented an opportunity to pause and consider their offensive Tweet will be less likely to create offensive Tweets.}

\begin{table*}[h!]
\centering
\caption{Offensive Replies, Total Replies, and Tweet Deletions per user by experiment assignment}
 \begin{tabular}{c|c c|c c|c c c c c}
 \toprule
 Metric & \multicolumn{2}{c}{Control} & \multicolumn{2}{c}{Treatment} & \multicolumn{5}{c}{Difference} \\
 \midrule
 {} & Mean & SD & Mean & SD & Abs. & \%  & SE& t & p (two-tailed)\\
 Offensive replies sent & 6.20 & 18.37 & 5.80 & 16.58 & -0.40 & -6.4\% & 0.07 & 5.33 & 9.6e-8 \\
 Total replies sent & 198.79 & 573.24 & 201.47 & 594.43 & 2.68 & 1.3\% & 2.50 & 1.08 & 0.28 \\
 Tweet deletions & 17.96 & 96.45 & 18.02 & 97.40 & 0.06 & 0.3\% & 0.41 & -0.15 & 0.88 \\
 \bottomrule
 \end{tabular}
\label{experiment_table}
  \small \\
  Note: nControl = 109,551; nTreatment = 109,501
\end{table*}

Interestingly, while we did observe offensive replies sent decrease, total replies sent did not move significantly and had a positive point estimate. This implies that the intervention can decrease a users' creation of offensive replies without affecting their participation in and non-offensive contributions to conversations.

Of the prompts delivered to treatment users, the following breakdown of actions were taken:
 \begin{itemize}
    \item 69\% of prompted Tweets were sent
    \item 9\% of prompted Tweets were cancelled
    \item 22\% of prompted Tweets were revised
\end{itemize} 

\begin{figure}[t]
\centering
\includegraphics[width=1.0\columnwidth]{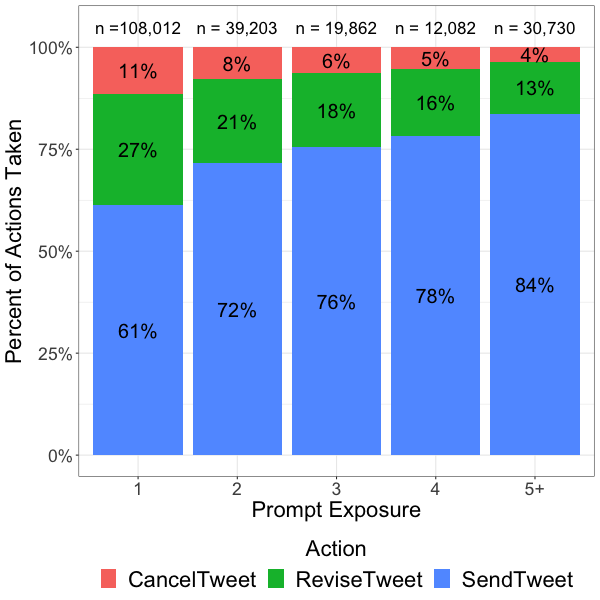}
\caption{Distribution of Actions Taken by Prompt Number}
\label{distribution_of_actions}
\end{figure}

When new features and interventions are introduced, it is common for novelty effects to occur where someone may react one way the first time seeing the novel feature and, after seeing this feature again, may not react the same way. To understand this, we looked at how the same users reacted to the intervention from their first exposure through multiple exposures. Users were most likely to revise or cancel their Tweet at the first exposure to the intervention. Upon receiving additional prompts, users became more likely to send their Tweet as-is, and less likely to revise or cancel. In Figure \ref{distribution_of_actions}, we show how the distribution of actions taken in response to a prompt changes from the first prompt exposure through to the fifth exposure, the proportion of those sending the Tweet increases as prompt exposure increases.

\begin{figure}[t]
\centering
\includegraphics[width=1.0\columnwidth]{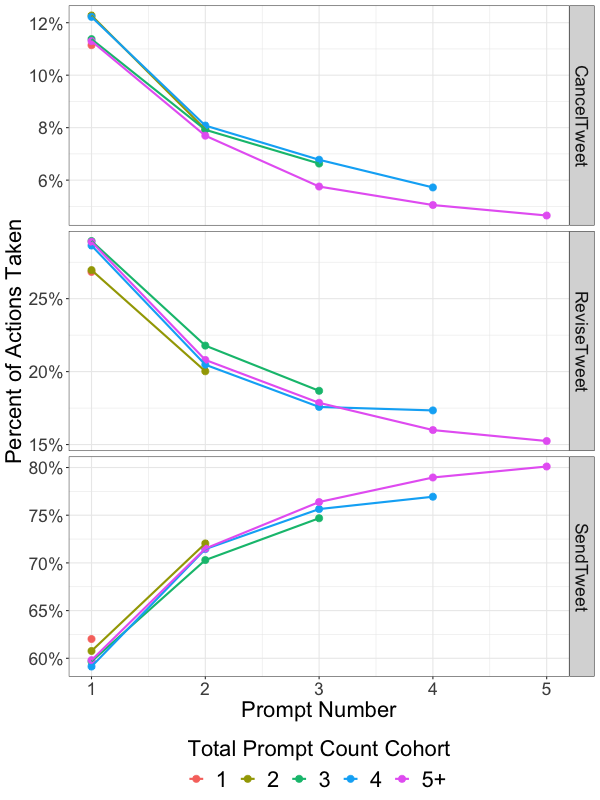}
\caption{Distribution of Actions Taken by Total Prompt Count Cohort and Prompt Number}
\label{distribution_of_actions2}
\end{figure}

To investigate this further, we wanted to see if there were differences in usage between individuals who were prompted just once throughout our experiment and those who were prompted multiple times. To understand how the distribution of actions taken differed across participants in our experiment, we split users into cohorts depending on how many total prompt-eligible Tweets they composed throughout the experiment. Conceptually, users in the “5+ Total Prompt Exposure” cohort would be more likely to intentionally and/or consistently create offensive content compared to those in the “1 Total Prompt Exposure” cohort and, perhaps, less likely to be affected by this intervention. Surprisingly, the send, revise, and cancel rates for the first prompt were comparable for all cohorts, seen in Figure \ref{distribution_of_actions2}. Regardless of how frequently someone creates offensive content, the rate of revision and cancellation were relatively similar on the first exposure of the prompt. Across these cohorts, the send rate increases with each subsequent prompt, suggesting that new forms of interventions can be effective even on users that routinely create offensive content.

\subsubsection{Revision Labeling Task}
To better understand the revisions resulting from users who chose to ``Edit'' their proposed Tweet, we sampled some of these edited Tweets and conducted a content analysis comparing the proposed text to the revised and posted text.

Labels were applied to describe the nature of the change between the proposed and posted text. For example if proposed Tweet text included the phrase ``fuck you'' but the posted Tweet text read ``f*** you'' this was labeled as ``Obfuscate'' to describe the use of asterisks to obfuscate profanity. Labeling was conducted by a single author to construct the codebook adding new labels as new types of edits emerged. This continued for approximately 40 Tweets until no new labels were being created. Given the relatively small number of ways that users edited their Tweets, the codebook only contains six distinct labels. The codebook - which includes a label, the description of the nature of the editing, a categorization as to whether the edit made the posted Tweet less/more/equally offensive, and example proposed and posted Tweet Text - is included in the Appendix (Table \ref{codebook}; Disclaimer: This table contains examples with offensive language.)

The labels from the initial set of 40 Tweets labeled while constructing the codebook were discarded and the codebook was applied by the single author who constructed the codebook to a set of randomly selected edited Tweets from the first week of the experiment. The same single coder applied labels until a saturation point was reached where the addition of 10 new labeled Tweets was not significantly changing the distribution of all labeled Tweets. This point was reached at 100 labeled Tweets. The same 100 Tweets were labeled blindly by the two other authors using the same codebook.

Intercoder reliability was measured on this dataset of 100 labeled Tweets by all three authors. All three coders had the same label 69\% of the time. Partial agreement - two of three coders having the same label - occurred 100\% of the time. Cohen's Kappa, used to measure agreement between two coders, was 0.801, 0.717, and 0.661. Fleiss' Kappa between all three was measured 0.725. Krippendorff's alpha was measured at 0.726. Overall this indicates substantial agreement within our labeled dataset.

The resulting data from labeling these edited Tweets showed that 37\% edited Tweets were revised to be less offensive than the originally proposed Tweet by removing profanity or other attacking language. 60\% of the edited Tweets labeled remained equally offensive as the proposed Tweet. Within this 60\%, the most common theme (23\% of all labeled Tweets) was making no change - where the proposed text was not modified in any substantive way before posting. The next most common theme (18\% of all labeled Tweets) was inserting asterisks or other substitutions to obfuscate a profane word. Lastly, while rare, there were 3\% of the edited Tweets labeled which were, in fact, edited to be more offensive than the proposed Tweet. Table \ref{codebook_distribution} includes the full distribution of labels in our dataset.

While this labeling exercise found that 37\% of edited Tweets were revised to be \emph{less} offensive than the originally proposed Tweets, some of these still remained offensive. Leveraging the same third-party platform used pre-experiment to asses the quality of the prompt eligibility criteria, annotators reviewed the posted Tweets which had been edited after being prompted in our experiment. For privacy reasons, the originally proposed text which triggered the prompt was not shown to annotators, and only the text of the public Tweet that was posted after editing was shown. In the same way previously described, annotators were shown 1,500 of the edited Tweets and asked ``Is this Reply Tweet toxic?'', with the options ``Yes''  and ``No''. 1,113 Tweets from this set were successfully labeled, and results from this indicated that 34\% of edited Tweets were labeled by the majority of annotators as not offensive.

These findings support H1 - \emph{People who are presented an opportunity to pause and consider their offensive Tweet will be less likely to create offensive Tweets.} In total, an estimated 8\% of prompted Tweets (37\% of the 22\% of edited Tweets) were revised to a less offensive alternative before being posted, while an estimated 7\% of prompted Tweets (34\% of the 22\% of edited Tweets) were revised to be unoffensive as determined by multiple third-party annotators.

Overall, the results of this experiment support the idea that presenting people with an opportunity to pause and consider their Tweet can reduce offensive content in conversations on Twitter. There was evidence to support all hypotheses except for H3.

\subsection{Hypothesis 2}
Our second hypothesis was aimed at understanding if the intervention could be used to shape people's behavior after a prompt is shown by discouraging people from creating similarly offensive Tweets in the future. Table \ref{table:2} shows how many people in control and treatment created varying amounts of prompt-eligible Tweets during the six week experiment. While control and treatment users composed their first prompt-eligible Tweet approximately the same number of times (as expected), treatment users were increasingly less likely to compose subsequent prompt-eligible Tweets. There were 4.3\% (p = 3.3e-15) fewer people in the treatment creating two prompt-eligible Tweets compared to the control. This decrease between control and test widens to 12.9\% (p = 1.8e-21) for individuals who created five prompt-eligible Tweets during the experiment. This suggests that the intervention may influence behavior beyond the moment a prompt is shown, and supports H2 - \emph{People who are presented an opportunity to pause and consider their offensive Tweet will be less likely to create other similarly offensive Tweets in the future.}

\begin{table}[h!]
\centering
\caption{Prompt-eligible Tweets composed by user assignment bucket}
 \resizebox{1.0\columnwidth}{!}{\begin{tabular}{c c c c c} 
 \toprule
 \makecell{Prompt\\Exposure\\Number} & Control & Treatment & \makecell{Treatment/\\Control} & \makecell{p\\(two-tailed)}\\
 \midrule
 1 & 109,446 & 109,394 & 100.0\% & 0.94\\ 
 2 & 41,532 & 39,731 & 95.7\% & 3.3e-15\\
 3 & 21,979 & 20,164 & 91.7\% & 1.4e-22\\
 4 & 13,885 & 12,294 & 88.5\% & 1.7e-25\\
 5 & 9,450 & 8,232 & 87.1\% & 1.8e-21\\
 \bottomrule
 \end{tabular}}
\label{table:2}
\end{table}

\subsection{Hypothesis 3}
Our next hypothesis, H3, was to understand if this intervention would result in fewer regrettable Tweets observed through a decrease in deletions after posting Tweets. As shown in Table \ref{experiment_table}, Tweet deletions did not move significantly. Thus, we did not find sufficient evidence for H3 - \emph{People  who  are  presented  an  opportunity to pause and consider their Tweet will be less likely to delete their Tweets after posting, as they are less likely to make regrettable Tweets that are hastily posted.}

\subsection{Hypothesis 4}
Our last hypothesis focused on understanding second-order effects of this intervention on the conversations surrounding prompted Tweets. As Seering et al (2017) found, people tend to mimic both pro and antisocial behaviors on social media. Given this, if we are able to successfully reduce some offensive content from being posted, we should expect a similar reduction in some of this mimicry behavior that would otherwise have occurred. To understand this, we looked at the number of replies made by people who were not in the experiment \emph{in response to} prompted Tweets made by people in the experiment. Of those replies in response to prompted Tweets, we calculated what fraction of them crossed the threshold for offensive content used in the analysis above. On average, 2.9\% of replies in response to treatment users' prompted Tweets were offensive replies, compared to 3.1\% for control users - a 5.8\% decrease (p = 0.02). This finding shows support for H4 - \emph{People who are presented an opportunity to pause and consider their offensive Tweet will receive fewer offensive replies to their Tweets.} Through reconsidering their own comments, users who are prompted can shape downstream conversations to becoming less offensive.

\section{Discussion}
The results of our experiment provide insight into how intervening during the time of Tweet creation by providing an opportunity for users to pause and reconsider their Tweet can decrease the use of harmful and offensive language on the platform:

\begin{itemize}
 \item \textbf{Prompting users to reconsider their Tweets can be an effective method for decreasing the number of potentially harmful and offensive Tweets posted.} While the majority of prompts resulted in users sending the Tweet as-is, 9\% of prompted Tweets were cancelled. In addition, 8\% of prompted Tweets were revised to have less offensive language. These revisions typically involved removal of profanity or other attacking language.
  
 \item \textbf{Interventions that decrease a user's offensive contributions to a conversation do not necessarily decrease non-offensive contributions.} We found a statistically significant decrease in offensive replies sent for prompted users, but not in total replies sent. This suggests that users receiving an intervention that reduces offensive contributions do not necessarily ``silence'' people or reduce their participation in conversations. In the public debate surrounding offensive content online, there is often a false dichotomy suggested between moderation and allowing speech freely. Our findings suggest that there are ways to encourage less offensive speech online without it coming at a cost to participation in online conversations. Given many users only occasionally post offensive content, perhaps it's unsurprising that many of them can be easily nudged to rephrase comments without offensive content. Evidence from this experiment shows that interventions can successfully focus on reducing contributions that contain offensive content.
 
 \item \textbf{The change in users' behavior persists beyond the immediate action taken upon receiving the prompt.} Beyond cancelling and revising their prompted Tweets, users assigned to treatment were less likely to compose prompt-eligible Tweets in the future. For example, 20\% fewer users assigned to treatment made five or more prompt-eligible Tweets during our experiment, compared to users assigned to control. This represents a broader and sustained change in user behavior, and implies that receiving prompts may help users be more cognizant of avoiding potentially offensive content as they post future Tweets. 
 
 \item \textbf{Novel interventions can be effective on both occasional and routine creators of offensive content.} We found that the rate of cancellations and revisions were similar for users, regardless of how many total prompt-eligible Tweets they composed throughout the experiment. For example, the cancellation rate for the first prompted Tweet was 11\% for both the cohort of users that received only one prompt throughout the experiment, and the cohort that received five or more prompts throughout the experiment. While there are certainly a small set of users who are intentionally creating offensive content, our findings suggest that even these users routinely creating offensive content may be open to reconsideration in response to a new intervention.
 
 \item \textbf{Interventions like this are subject to novelty effects} While the the rate of cancellation was 11\% for those first exposed to the intervention, this rate decreased for each additional exposure to the intervention. This may be partially attributed to the behavior change mentioned above, where prompted users are already reconsidering potentially harmful and offensive context prior to sending a Tweet. But this finding also highlights the need to change these interventions over time with updated copy, design, or interactions to ensure persistent efficacy.
 
 \item \textbf{Reducing the number of harmful and offensive Tweets that a user sends also reduces the number that they receive.} On online platforms, offensive content can beget more offensive content. When a user turns a conversation negative, or when they extend an already negative thread, their reply Tweet can become the new root for other users to contribute negatively. By reducing the number of offensive Tweets, this intervention also reduced the number of opportunities for other users to respond with additional offensive Tweets. This can have a doubly beneficial effect, where the prompted user is both less likely to be a target for offensive content and less likely to be a spreader of offensive content. These findings demonstrate the significant impact that these upstream, ex ante, interventions can have. Compared to the more traditional ex post moderation approach, these ex ante moderation strategies benefit from downstream and network effects resulting when offensive content is avoided altogether.
\end{itemize}

One of the biggest limitations to the findings and claims in this study is that the impact of this intervention is ultimately dependent on the method used to identify and determine which Tweets are presented with a prompt. Whatever results observed during this experiment, it should be expected that results will vary depending on what types of offensive content is provided with such an intervention and the effectiveness or accuracy of your method for identifying such content. That said, the adoption of such a similar intervention across multiple social media platforms referenced above indicates that this approach appears to have viability on some portion of offensive content in different contexts online. Furthermore, while this intervention had modest success in reducing some amount of offensive content, it should be noted that, of course, it does not ``solve'' the problem of offensive content online. As discussed earlier, given the breadth of motivations behind offensive content creation, this intervention should only be expected to appeal to some creators while other approaches must be used for the remainder.

Another limitation to this study is that it was conducted on English Tweets only. Depending on how well a similar algorithm performs at identifying harmful content across languages, the results we saw from this study may not generalize to non-English Tweets. Additionally, and more importantly, it is expected that users from different cultural contexts may react differently when prompted to reconsider their offensive Tweets.

Future research can build on these findings to further our understanding of offensive content online and how to make interventions more effective. First, our work was largely quantitative, and did not address the motivations for why users might send, revise, or cancel offensive Tweets. Qualitative studies could help uncover the motivations for why users responded to the prompts as they did, and potentially uncover subsets of users or contexts where prompts are more or less effective. Research can also investigate how users' perceptions of offensive content on Twitter as well as perceptions of their own behavior change after receiving interventions. This might help to further explain the mechanisms through which this intervention is proving successful. While we found that prompted users were less likely to contribute harmfully to a thread, we do not know how the experience of being prompted affected their own experience on Twitter. For example, if they felt their response was in self-defense to another offensive Tweet, they might see the prompt as unjustified and a blocker to defending themselves on the platform. Related, although a rare occurrence, we did observe that some individuals prompted chose to edit their Tweet to be more offensive. Qualitative research could help to uncover why some people react in such a way shedding light on improvements that could be made to the intervention to avoid these instances.

Another promising path for future inquiry is related to the framing and messaging techniques used in this intervention. In our experiment, we had a control who saw no prompt and a single test group which did see a prompt. At a minimum, there is a lot of opportunity to continue testing the messaging to understand which frame might prove most successful. Additionally, there is significant research related to persuasion and message framing that would suggest that varying message frames would perform differently across populations depending on culture, identity, and many other factors \cite{feinberg2019moral, cialdini2009influence}. While some individuals may be persuaded away from posting offensive content with the threat of punishment, others may respond to descriptive norms.

Finally, there is an opportunity for future research on the topic of regrettable Tweets and deletion patterns. Although our experiment found that prompting users reduces offensive replies sent, it did not find a relationship between prompting and Tweet deletion. Exploring what types of Tweets are self-moderated can help us understand how users recognize and regret their own behavior. This might help inform a more targeted intervention aimed at helping people avoid those self-moderated Tweets.

\section{Conclusion}

Abusive behavior and offensive content on social media continues to be a problem that every platform is actively working to manage. Much of the previous attention has been paid towards: techniques and approaches to build machine learning algorithms to detect this content more quickly; finding more efficient ways to report, review, and remove offensive content; and providing more transparent and fair processes for those who have content removed. This experiment adds to a growing body of evidence demonstrating the opportunity to move interventions further upstream, providing people an opportunity to reflect on their behaviors online and discourage offensive content from being posted altogether. 

We found through an online A/B experiment that prompting people to reconsider their Tweet as they are posting offensive content changes their behavior both at posting time and in their future postings. People who were prompted to reconsider their Tweet cancelled it 9\% of the time and revised it 22\% of the time (37\% of these was to a less offensive alternative). Overall, these users posted 6\% fewer offensive Tweets and saw a 6\% decrease in the proportion of offensive replies to their prompt-eligible Tweets.

Future work aims to better understand the motivations behind some of these offensive Tweets so that more targeted interventions can be created to further reduce offensive and harmful content online.

\section{ Acknowledgments}

The authors would like to thank the following for the contribution to this project and feedback on the manuscript: Alberto Parrella, Stefan Wojcik, Shaili Jain, Charis Redmond, Cody Elam, Allen Lee, and the rest of the Incentives team at Twitter. This research was funded by Twitter, Inc. At the time the research was conducted, all three authors were employed by Twitter, Inc.

\bibliography{main}
\bibliographystyle{aaai21}

\begin{table*}
\centering{\large{\textbf{Appendix}}}
\caption{Codebook used in content analysis to compare proposed and posted Tweet text. Fictitious examples provided with bolded text to indicate changes.}
\centering
 \begin{tabular}{|c|c|L{4cm}|L{3.25cm}|L{3.25cm}|}
 \toprule
 Label & Change & Description & Example Proposed Text & Example Posted Text  \\
 \midrule
 Removed Profanity & Less Offensive & Profanity or some other attacking language is removed from proposed text & Damn. What a moron. You wanna go out \textbf{being cucked by trump}?
 
 & Damn. What a moron. You wanna go out \textbf{like that} ? 
 
 \\
 Obfuscate & Equally Offensive & Adding an asterisk, removing a letter, changing to phonetic spellings, replacing with an emoji or some other obfuscation is used on profane words or phrases
 
 & Sam is a piece of \textbf{shit}. & Sam is a piece of \textbf{\emojipoop}.\\ 
 Abbreviation & Equally Offensive & Similar to ``Obfuscate'' but limited only to posts in which users abbreviate or shorten a profane word or phrase
 
 & You are the one supporting them so what \textbf{the fuck} is wrong WITH YOU? & You are the one supporting them so what \textbf{tf} is wrong WITH YOU?
 
 \\
 Changed Attack & Equally Offensive & A revision was made to replace one attack with another equally offensive attack
 
 & Fuck the word of god. Jesus Christ is \textbf{a bitch} & Fuck the word of god. Jesus Christ is \textbf{not real. Made up and gay} \\
 No Change & Equally Offensive & No substantive change was made to the offensive part of the proposed Tweet text
 
 & You \textbf{rediculous}\emojiclown wtf is wrong with you? \emojiwavinghand & You\emojiclown wtf is wrong with you? \emojiwavinghand
 
 \\
 Added Attack & More Offensive & An attack was added to the proposed Tweet text & I don’t need your bitch ass & I don’t need your bitch ass. \textbf{you have no friends you loser}.
 
 \\
 \bottomrule
 \end{tabular}
\label{codebook}
\end{table*}

\begin{table*}
\centering
\caption{Distribution of labeled categories for the 100 edited Tweets labeled.}
 \begin{tabular}{|p{3cm}|c|c|c|c|}
 \toprule
 Label & Coder 1 & Coder 2 & Coder 3 & Majority Label \\
 \midrule
 Removed Profanity
 
 & 36\% & 37\% & 28\% & 37\% \\
 Obfuscate
 
 & 19\% & 17\% & 17\% & 18\%\\
 Abbreviation
 
 & 9\% & 10\% & 10\% & 9\%\\
 Changed Attack
 
 & 11\% & 4\% & 13\% & 9\%\\
 No Change
 
 & 23\% & 29\% & 29\% & 23\%\\
 Added Attack
 
 & 2\% & 3\% & 3\% & 3\%\\
 \bottomrule
 \end{tabular}
\label{codebook_distribution}
\end{table*}

\end{document}